\documentclass[12pt]{article}

 \def\lddots{\mathinner{\mkern1mu\raise1pt\hbox{.}\mkern2mu

\raise4pt\hbox{.}\mkern2mu\raise7pt\vbox{\kern7pt\hbox{.}}\mkern1mu}}

\makeatletter

\def\numberbysection{\@addtoreset{equation}{section}
 \def\theequation{\thesection.\arabic{equation}}}

\makeatother

\numberbysection

\newcommand{\be}{\begin{eqnarray}}
\newcommand{\ee}{\end{eqnarray}}
\newcommand{\non}{\nonumber}

\newcommand{\tr}{\mathop{\rm tr}\nolimits}

\begin{document}

\begin{titlepage}

\strut\hfill{LAPTH-973/03}

\vspace{0.5in}

\begin{center}

\LARGE  Fused integrable lattice models with quantum impurities and open boundaries \\[1.0in]

\vspace{.3in}

\large{Anastasia Doikou} \footnote{e-mail: doikou@lapp.in2p3.fr} \\

\vspace{.3in}

\normalsize{Theoretical Physics Laboratory of Annecy--Le--Vieux,\\
LAPTH, B.P. 110, Annecy--Le--Vieux, F-74941, France}\\

\end{center}

\vspace{1in}

\begin{abstract}
The alternating integrable spin chain and the $RSOS(q_{1},q_{2};p)$ model in
the presence of a quantum impurity are investigated. The boundary
free energy due to the impurity is derived, the ratios of the
corresponding $g$ functions at low and high temperature are
specified and their relevance to boundary flows in unitary minimal
and generalized coset models is discussed. Finally, the
alternating spin chain with diagonal and non--diagonal integrable
boundaries is studied, and the corresponding boundary free energy
and $g$ functions are derived.

\end{abstract}

\end{titlepage}

\section{Introduction}

Two dimensional exactly solvable models with boundaries have
attracted a great deal of research interest recently from the
point of view of boundary conformal field theory \cite{cardy}, and
critical behavior \cite{af}, but also because of the rich variety
of physical phenomena they display, which in principle can be
exactly investigated (see e.g. \cite{cherednik}--\cite{DVGR3}).
There has been also much interest on problems related to quantum
impurities, mainly because of the important role they play in low
dimensional physics \cite{lut}, but also because of their
relevance to boundary conformal field theory \cite{cardy, af,
lesage, ahnrim}. There are numerous studies related to quantum
impurities yielding a great number of interesting and useful
results (see e.g. \cite{ftw, ande, wang}). In this article we
focus basically on the thermodynamic analysis of lattice
integrable systems in the presence of quantum impurities and
integrable open boundaries.

For both integrable lattice models and relativistic integrable
field theories, in the bulk, the corresponding free energy has been
derived and the conformal properties have been extensively studied
\cite{yy}--\cite{dd} by means of the thermodynamic Bethe ansatz.
It is however of great interest to extend these studies for
integrable models with boundaries. In analogy to the bulk case,
when boundaries are added, the corresponding boundary free energy
and the so called $g$ function, which characterizes the
ground state degeneracy due to the boundaries,
 \cite{af, lmss, fsw, drtw, alcaraz, desa} can be specified  by means of the
thermodynamic Bethe ansatz. On the other hand statistical systems
at the critical point it is known to display conformal invariance
\cite{pol, bpz}, therefore they can be associated with certain
conformal field theories. The low temperature behavior of the free
energy per unit length of such system, in the bulk, is described by
\cite{bc, af1} \be f(T) = f_{0} - {\pi c \over 6 u}T^{2}+\dots,
~~T\ll 1, \ee where $c$ is the central charge of the effective
conformal field theory. Furthermore, when boundaries are added the obtained free energy is modified up to an ${1 \over L}$ contribution ($L$ denotes the size of the system) namely,
\cite{af} \be f(T)= f_{0}(T) -{T\over L} \ln g, \label{gfunction} \ee
where the ground state degeneracy $g$ is expected to be related
with the boundaries of the system.
One of the main aims, when studying such systems is to employ proper techniques in order to specify the central charge and the ground state degeneracy ---when boundaries are present.

As already mentioned from the integrable systems point of view,
the central charge and the ground state degeneracy $g$ can be
identified by employing thermodynamic Bethe ansatz techniques (see
e.g. \cite{yy}--\cite{dd}, \cite{lesage, ahnrim, lmss, fsw, drtw,
desa}). In this study in particular the alternating integrable spin chain
\cite{VEWO} and the $RSOS(q_{1},q_{2};p)$ model \cite{doikou} in
the presence of a quantum impurity (``Kondo type'' boundaries see
e.g. \cite{ftw, ande}) are investigated via the thermodynamic
Bethe ansatz, and the corresponding free energy is derived at low
and high temperature. The relevance of the results for the
$RSOS(q_{1},q_{2};p)$ model to the boundary flows in minimal
\cite{lesage} and generalized coset models \cite{ahnrim} is
discussed. Finally the alternating open spin chain with diagonal
and non--diagonal integrable boundaries is considered and the
corresponding boundary free energy and the $g$ functions for the
left and right boundaries are determined by first principle
calculations at low and high temperature.

\section{Quantum Impurity}

\subsection{The alternating spin chain}

Let us first focus on the alternating spin chain \cite{VEWO} in
the presence of a quantum impurity. For what follows it is
necessary to introduce the basic constructing element of the
model, namely the $R$ matrix, which is a solution of the
Yang--Baxter equation \cite{baxter, korepin} \be
R_{12}(\lambda_{1} - \lambda_{2})\ R_{1 3}(\lambda_{1})\
R_{23}(\lambda_{2}) = R_{23}(\lambda_{2})\ R_{1 3}(\lambda_{1})\
R_{1 2}(\lambda_{1} - \lambda_{2})\,. \label{YBE} \ee We consider
the $R$ matrix obtained in \cite{KR}, namely
 \be R_{0k}^{1,s} =
\sinh \mu \Big (\lambda + i ({1\over 2} +\sigma^{3} \otimes S^{3}) \Big) + \sinh \mu \Big (\sigma^{+}\otimes S^{-} +\sigma^{-} \otimes S^{+}\Big )
                                                       \label{Rmatrix1}   \ee
where
$S^{3}$, $S^{\pm}$  act in general, on a $2S+1$ dimensional space
$V=C^{2S+1}$, and they satisfy the following commutation
relations \be \left[ S^+ \,, S^-\right] &=&  {\sinh 2 i \mu S^3 \over \sinh i \mu } \,, ~~\left[ S^3
\,, S^\pm \right] = \pm S^\pm , \non\\ S^{3}|0 \rangle &=& S|0 \rangle ,~~S^{+}|0 \rangle =0. \ee
 We can now define the transfer matrix of the
chain \footnote{If we derive the transfer matrix of the model with
proper inhomogeneities we obtain massless relativistic dispersion
relations for the particle--like excitations of the model
\cite{bado}.} \be t= tr_{0} T_{0}(\lambda) \ee where \be
T_{0}(\lambda) = R_{0 2N+1}^{q}(\lambda-\Theta)R_{0 2N}^{1}(\lambda)  R_{0
2N-1}^{2}(\lambda)\cdots R_{0 2}^{1}(\lambda)  R_{0 1
}^{2}(\lambda) \,, \label{mono2} \ee and $R^{i}$ is related to the
spin $S_{i}={q_{i} \over 2}$ ($i=1,2$) representation ($R^{q}$ is related to the
spin ${q\over 2}$ representation) (\ref{Rmatrix1}). Following the
standard Bethe ansatz method described in e.g. \cite{VEWO}, \cite{FT2}--\cite{bado}, the Bethe equations are obtained
\begin{equation}
e_{q_{1}}(\lambda_{\alpha})^{N}e_{q_{2}}(\lambda_{\alpha})^{N}e_{q}(\lambda_{\alpha}-\Theta)=
-\prod_{\beta=1}^{M} e_2(\lambda_{\alpha}-\lambda_{\beta})
\label{BEI}
\end{equation}
where
\begin{equation}
e_{n}(\lambda; \nu)=\frac{\sinh \mu(\lambda+{in\over 2})}{\sinh
\mu(\lambda-{in \over 2})}, ~~\nu={\pi \over \mu}. \nonumber
\end{equation}
Notice that the main difference between the usual bulk case \cite{VEWO} (without impurities) and (\ref{BEI})
is the appearance of the $e_{q}$ term in the left hand side of
(\ref{BEI}) due to the presence of the spin ${q \over 2}$
impurity. Moreover we should mention that $q_{1}$, $q_{2}-q_{1}$
play the role of ``flavors'' in accordance with the picture in \cite{ande}, where the spin ${f\over 2}$ ($f$ flavor) chain
is studied with impurity of spin $s$.

 In the thermodynamic limit, $N \to \infty$, the string hypothesis is valid \cite{g, t1, FT2}, namely
the solutions of equation (\ref {BEI}) can be grouped into strings of
length $n$ with the same real part and equidistant imaginary parts
\be \lambda_{\alpha}^{(n,j)} &=&\lambda_{\alpha}^n + {i\over
2}(n+1-2j),~~j=1,2,...,n, \non\\
\lambda_{\alpha}^{0,s}&=&\lambda_{\alpha}^{0}+i\frac{\pi}{2\mu},
 \label{STR} \ee where $\lambda_{\alpha}^n$ is real, and
$\lambda^{0,s}$ is the negative parity string and $\lambda^{0}$ is
real. In order to formulate the thermodynamic Bethe ansatz we consider all the strings with length $n=1,\ldots ,\nu-1$
plus the negative parity string.
The Bethe ansatz equations (\ref{BEI}) can be written after we
apply the string hypothesis \cite{g, t1} \be
X_{nq}(\lambda_{\alpha}^n-\Theta)\prod _{j=
1}^{2}X_{nq_{j}}(\lambda_{\alpha}^n)^{N} =
(-)^{n}\prod_{m= 1}^{\nu -1} \prod_{\beta=1}^{M_{m}}E_{nm}(\lambda_{\alpha}^{n}-
\lambda_{\beta}^{m})
\prod_{\beta=1}^{M_{0}}G_{n1}(\lambda_{\alpha}^{n}-\lambda_{\beta}^{0})
 \label{k0} \ee
\be
g_{q}(\lambda_{\alpha}^0-\Theta)\prod _{j=
1}^{2}g_{q_{j}}(\lambda_{\alpha}^0)^{N} = -\prod_{m=
1}^{\nu -1} \prod_{\beta=1}^{M_{m}}
G_{1m}(\lambda_{\alpha}^{0}-\lambda_{\beta}^{m})\prod_{\beta=1}^{M_{0}}
e_{2}(\lambda_{\alpha}^{0}-\lambda_{\beta}^{0}), \label{K2} \ee where
\begin{equation}
g_{n}(\lambda;\nu) = e_{n}(\lambda \pm {i \pi\over 2 \mu})={\cosh
\mu(\lambda +{in\over 2}) \over \cosh \mu(\lambda -{in\over 2})}\\,
\label{gg}
\end{equation}
and $X_{nm}$, $E_{nm}$, and $G_{nm}$ are given in the appendix (\ref{ap1}).

We introduce the  densities of the holes $\tilde \rho_{n}$ and pseudo--particles
$\rho_{n}$, and once we take the logarithm and the derivative of the Bethe ansatz equations (\ref{k0}), (\ref{K2}) we conclude \be  \tilde \rho_{n}(\lambda)=
{1\over 2}( Z_{nq_{1}}(\lambda)+ Z_{nq_{2}}(\lambda)) +
{1\over L} Z_{nq}(\lambda-\Theta)
-\sum_{m=1}^{\nu -1}A_{nm}
* \rho_{m}(\lambda) - B_{1n} * \rho_{0}(\lambda) \non\\
 -(\rho_{0}(\lambda) + \tilde \rho_{0}(\lambda)) =  {1\over 2}(b_{q_{1}}(\lambda)
+b_{q_{2}}(\lambda)) +{1\over L}b_{q}(\lambda-\Theta)
 -\sum_{m=1}^{\nu -1}B_{1m} * \rho_{m}(\lambda)
-a_{2}*\rho_{0}(\lambda) \label{densitiesi} \ee where $\rho_{0}$
is the density of the negative parity string, and $L=2N$ is the
length of the spin chain\footnote{We treat the impurity, which
sits at the $2N+1$ site of the chain, separately, therefore $L=2N$
is the length of the bulk part.}, and the quantities $Z_{nm}$,
$A_{nm}$, $B_{nm}$, and $b_{n}$ are given in the appendix
(\ref{Z}), (\ref{A}), (\ref{b}), and (\ref{ro0}). It is also
convenient to solve $\rho_{n}(\lambda)$ in terms of $\tilde
\rho_{n}(\lambda)$, therefore we consider the convolution of the
first of the equations (\ref{densitiesi}) with the inverse of
$A_{nm}$ \be \hat A_{nm}^{-1} = \delta_{nm} -\hat
s(\omega)(\delta_{nm+1} +\delta_{nm-1}) \ee where $\hat s(\omega)
= {1\over 2 \cosh{\omega \over 2}}$, $s(\lambda) = {1\over 2
\cosh\pi \lambda}$.
 Having in mind the identities
\be A_{nm}^{-1} * Z_{mq_{i}}(\lambda) = s(\lambda)
\delta_{nq_{i}},  ~~ A_{nm}^{-1} * B_{1m}(\lambda) = - s(\lambda) \delta_{n
\nu -2} \label{ident} \ee  we obtain the following expressions  \be
\rho_{n}(\lambda) &=& {1 \over 2} s(\lambda)
(\delta_{nq_{1}}+\delta_{nq_{2}}) - \sum_{m=1}^{\nu -1}
A_{nm}^{-1}* \tilde \rho_{m}(\lambda) -\delta_{n \nu -2} s *\tilde
\rho_{0}(\lambda) \non\\ &+& \delta_{n \nu -2} s *s *
\tilde \rho_{0}(\lambda) +{1\over L}s(\lambda- \Theta) \delta_{nq} \non\\
\rho_{0}(\lambda) &=&- \tilde \rho_{0}(\lambda) +s*\tilde
\rho_{\nu -2}(\lambda) \label{densitiesim} \ee note the extra
${1\over L}$ contribution term in the first of the two above
equations, which is due to the impurity. Let us now derive the free energy of
the system $f=e-Ts$, which is given by
\be
f= e-Ts &=& -
{1\over 2}\sum_{n=1}^{\nu -1} \int_{-\infty}^{\infty}d\lambda
(Z_{nq_{1}}(\lambda) +Z_{nq_{2}}(\lambda))\rho_{n}(\lambda)-{1\over 2} \int_{-\infty}^{\infty}d\lambda (b_{q_{1}}(\lambda)
+b_{q_{2}}(\lambda))\rho_{0}(\lambda)\non \\&-& T\sum_{n=1}^{\nu -1}
\int_{-\infty}^{\infty} d \lambda \Big (\rho_{n}(\lambda)
\ln(1+{\tilde \rho_{n}(\lambda) \over \rho_{n}(\lambda)}) + \tilde
\rho_{n}(\lambda) \ln(1+{
\rho_{n}(\lambda) \over \tilde \rho_{n}(\lambda)})\Big ) \non\\
&-& T\int_{-\infty}^{\infty}d \lambda \Big (\rho_{0}(\lambda)
\ln(1+{\tilde \rho_{0}(\lambda) \over \rho_{0}(\lambda)}) + \tilde
\rho_{0}(\lambda) \ln(1+{ \rho_{0}(\lambda) \over \tilde
\rho_{0}(\lambda)})\Big ). \label{entropy}\ee
The thermodynamic Bethe ansatz equations are obtained by minimizing the free energy ($\delta f =0$) and by virtue of (\ref{densitiesi}), we conclude that they coincide with the ones of the
model without impurities \cite{bydo} and they are
given by  \be T \ln \Big
(1+\eta_{n}(\lambda)\Big ) &=& -{1\over
2}(Z_{nq_{1}}(\lambda)+Z_{nq_{2}}(\lambda)) + T\sum_{m=1}^{\nu
-1} A_{nm}* \ln \Big (1+\eta_{m}^{-1}(\lambda)\Big ) \non\\
&-& T B_{1n}*\ln \Big (1+\eta_{0}^{-1}(\lambda)\Big ) \non\\ T \ln
\Big ({1+\eta_{0}(\lambda) \over 1+\eta_{0}^{-1}(\lambda)}\Big )
&=& -{1 \over 2}(b_{q_{1}}(\lambda)+b_{q_{2}}(\lambda)) +T
\sum_{m=1}^{\nu -1} B_{1m}* \ln \Big (1+\eta_{m}^{-1}(\lambda)\Big
) \non\\ &-& T a_{2}*\ln \Big (1+\eta_{0}^{-1}(\lambda)\Big ),
\label{TBA} \ee
where $\eta_{n} = {\rho_{n} \over \tilde \rho_{n}}$.
 Alternatively we can write the thermodynamic Bethe ansatz equations
by virtue of (\ref{densitiesim}) in the following form, $\eta_{n}(\lambda) =e^{{\epsilon_{n}(\lambda)
\over T}}$, \be \epsilon_{n}(\lambda)
&=&s(\lambda)*T \ln (1+ \eta_{n+1}(\lambda))(1+
\eta_{n-1}(\lambda)) -{1\over 2}s(\lambda)(\delta_{nq_{1}}
+\delta_{nq_{2}}) \non\\ &+& \delta_{n \nu
-2}s(\lambda)*T \ln \Big( 1+ \eta_{0}^{-1}(\lambda) \Big ) \non\\
\epsilon_{\nu -1}(\lambda) &=& s(\lambda)*T \ln (1+ \eta_{\nu
-2}(\lambda)) \non\\ \epsilon_{0}(\lambda) &=& - s(\lambda)*T \ln
(1+ \eta_{\nu -2}(\lambda)). \label{TBA2} \ee Although the
thermodynamic Bethe ansatz equations remain the same as in the
bulk case, the free energy of the model is modified because of the
presence of the quantum impurity. In particular, there is a non
trivial contribution to the free energy which is given by the
following expression, after we apply (\ref{densitiesim}) and
(\ref{TBA}), (\ref{TBA2}) to (\ref{entropy}) \be f = e_{0} + f_{0}
+f_{b} +O({1\over L}) \label{contribution} \ee where $f_{0}$ is
the bulk free energy given by \be f_{0} = -{T \over
2}\int_{-\infty}^{\infty}d\lambda
s(\lambda)\ln(1+\eta_{q_{1}}(\lambda))(1+\eta_{q_{2}}(\lambda)).
\label{fe0} \ee $e_{0}$ is the energy of the the state with the
seas of strings with length $q_{1}$, $q_{2}$ filled  \be e_{0} =
-{1\over 4} \sum_{i,j =1}^{2}\int_{-\infty}^{\infty}d\lambda
Z_{q{i}q_{j}}(\lambda) s(\lambda) -{1\over 2L} \sum_{j
=1}^{2}\int_{-\infty}^{\infty}d\lambda Z_{qq_{j}}(\lambda-\Theta)
s(\lambda) \label{energyim} \ee where the non--trivial ${1\over
L}$ contribution to the bulk ground state energy is due to the
impurity. Finally, $f_{b}$ is the free energy contribution of the
impurity (after we make the shift $\lambda \to \lambda -{1\over
\pi} \ln T $) \be f_{b} = -{ T \over L} \int_{-\infty}^{\infty}
d\lambda s(\lambda-{1\over \pi} \ln T ) \ln (1+\eta_{q}(\lambda)).
\label{freeboundary1}\ee We can accurately evaluate differences of
boundary free energies, or ratios of $g$ functions
(\ref{gfunction}), for different temperatures and not specific
values at each temperature. This happens basically because an
overall ${1\over L}$ contribution, which can not be explicitly
evaluated, may survive in the bulk calculation (see
(\ref{contribution})) of the free energy as well (see also
\cite{lmss, fsw, drtw}). The boundary free energy, and the
corresponding $g$--function (\ref{gfunction}) can be computed at
any temperature by employing numerical methods, however it is
possible to make analytical calculations at $T \to \infty$
``weak--coupling'' and $T \to 0$ ``strong--coupling'' point (see
e.g. \cite{af, ande}).

In particular, for $T \to \infty $ the main contribution to the
integral in (\ref{freeboundary1}) comes from the $\lambda \to
\infty $ behavior. The corresponding behavior of
$\eta_{q}(\lambda)$ is given by
 (see also \cite{BT, bydo}) \be (1+\eta_{n}^{\infty})^{-1} &=&{1\over(n+1)^{2}},
~~n=1,\ldots \nu-2 \non\\ (1+\eta_{\nu -1}^{\infty})^{-1} &=&{1\over \nu},
~~(1+\eta_{0}^{\infty})^{-1} = 1- {1\over \nu}, \label{sol1} \ee
which obviously does not depend on
$q_{1}$, $q_{2}$. The boundary free energy contribution is for
$q < \nu -1$ ``weak--coupling'' point (see e.g. \cite{af, ande}),
\be f_{b}^{\infty} = -{T \over L}
\ln(q+1). \label{g} \ee
In the special value where $q =\nu
-1$ the boundary energy becomes following (\ref{sol1}) \be f_{b}^{\infty} =
-{T\over 2L} \ln(\nu),  \ee which is the half  of the expected value.
In the isotropic case $\nu \to \infty$, $q$ can take any value
from 1 to $\infty$ and the impurity contribution is given by
(\ref{g}).

When $T \to 0$ the main contribution to the integral in (\ref{freeboundary1}) comes
from the $\lambda \to - \infty$ behavior --- recall that we considered the shift
$\lambda \to \lambda -{1\over \pi} \ln T $. The quantity $1+\eta_{n}$ for $\lambda
\to - \infty$ and $T \to 0$ is given by
\cite{BT, bydo} \be (1+\eta_{n}^{0})^{-1} &=& {\sin
^{2}({\pi \over q_{1} +2}) \over \sin ^{2}({\pi (n+1) \over q_{1}
+2})}, ~~n=1, \ldots ,q_{1}-1, ~~(1+\eta_{q_{1}}^{0}) =1 \non\\
(1+\eta_{n}^{0})^{-1} &=& {\sin ^{2}({\pi \over q_{2} - q_{1} + 2}) \over
\sin ^{2}({\pi (n-q_{1}+1) \over q_{2}-q_{1} +2})},
~~n=q_{1}+1, \ldots ,q_{2}-1, ~~(1+\eta_{q_{2}}^{0}) =1 \non\\
(1+\eta_{n}^{0})^{-1}
&=&{1 \over (n-q_{2}+1)^{2}}, ~~n=q_{2}+1, \ldots ,\nu-2 \non\\
(1+\eta_{\nu-1}^{0})^{-1} &=&{1\over \tilde \nu}, ~~(1+\eta_{0}^{0})^{-1} = 1- {1\over
\tilde \nu}. \label{sol2} \ee where $\tilde \nu = \nu -q_{2}$.
The situation is more complicated now, because the behavior of $\eta_{q}(\lambda)$ depends clearly on the flavors (see (\ref{sol2})).
In particular, for $q=q_{i}$ the boundary entropy is $f_{b} \propto T^{2}$,
and
this is the
completely screened case (see e.g \cite{ande}).
For $q < q_{1}$
\be
f_{b}^{0} = - {T \over L} \ln {\sin \Big ({\pi(q+1)\over q_{1}+2} \Big) \over \sin
\Big ({\pi \over q_{1}+2} \Big)} \label{f10} \ee
this is the behavior of the non-trivial ``strong--coupling'' point
(see also \cite{af, ande}) with flavor $q_{2}$ and impurity spin $q$. For $q_{1} < q < q_{2}$
\be
f_{b}^{0} = -{T \over L} \ln {\sin \Big ({\pi(q-q_{1}+1)\over q_{2}-q_{1}+2} \Big) \over \sin
\Big ({\pi \over q_{2}-q_{1}+2} \Big)}, \label{f20} \ee
this case also corresponds to a ``strong--coupling'' point, with flavor $q_{1}-q_{2}$
and a reduced spin impurity $q-q_{2}$.
For $q>q_{2}$
\be
f_{b}^{0}= -{T \over L}\ln(q-q_{2}-1) \label{f3} \ee this is the partially screened case with
 reduced impurity spin $q-q_{1}$. Finally, for $q=\nu-1$ (\ref{sol2})
\be f_{b}^{0}=-{T \over 2L} \ln (\nu -q_{2}). \label{f4} \ee
We should emphasize again  that the boundary free energy is calculated up to an overall
${1\over L}$ contribution which we are not able to derive explicitly. Therefore, we
consider only differences of the free energy for different temperatures, and from that
via relation (\ref{gfunction}) we deduce the ratio of the $g$ function,
\be
\ln{g_{\infty} \over g_{0}} =  {1\over 2}\ln{(1+\eta_{q}^{\infty})\over
(1+\eta_{q}^{0})} \label{result} \ee where $\ln (1+\eta_{q}^{\infty, 0})$
is derived in  (\ref{sol1}), (\ref{sol2}).
The model under study is related to $WZW_{\delta q}$ and
$WZW_{q_{2}}$ model, as already concluded  in \cite{AM, bydo}.
Indeed the central charge conjectured in \cite{AM}, and found in \cite{bydo} is given
by
\be c={3q_{1} \over
q_{1} +2} +{3 \delta q \over \delta q +2} \label{central} \ee and
it is expressed as the sum of the central
charges of the $SU(2)$ $WZW$ models at level $q_{2}$ and $\delta
q$ ($\delta q =q_{2} -q_{1}$). Therefore we expect that our results (\ref{g})--(\ref{f4}), (\ref{result})
should be related to
boundary flows in $WZW_{k}$ models in analogy to the bulk case.
Finally, we should note that our results for the free energy (\ref{g})--(\ref{f4})
for $q_{1} =q_{2}$ coincide with the ones found in \cite{ande} for the spin ${f\over 2} ={q_{1} \over 2}$
chain with spin $s={q \over 2}$ impurity.

In general the impurity sitting at the $2N+1$ site of the chain can
be thought as an immobile ``particle'' with constant rapidity
$\Theta$. Therefore, the particle--like excitations of the chain
can interact (scatter) with the impurity giving rise to specific
scattering amplitudes see e.g. \cite{fendleykink}, which we are
going to study in detail elsewhere. The scattering should involve
in addition to the usual $XXX$ part (in the isotropic limit of our
model), $RSOS$ type scattering as well \cite{fendleykink}. This is expected since our
chain is related to $WZW_{k}$ model,
and it is known (see e.g. \cite{Bernard}) that
the $S$--matrix that describes $WZW_{k}$ models has apart from the
$SU(2)$ invariant part an $RSOS$ part as well, namely
\cite{Bernard} \be S_{WZW_{k}} = S_{SU(2)} \otimes S_{RSOS}^{(k)}.
\label{sw} \ee $S_{SU(2)}$ is the usual $XXX$ $S$ matrix
and $S_{RSOS}^{(k)}$ is the $S$ matrix for the $RSOS$ model with
restriction parameter $k+2$ see also \cite{resrsos}.

\subsection{The generalized $RSOS(q_{1}, q_{2}; p)$ model}

It is known that the effective conformal field theory for the critical $RSOS(1,1)$
model is the unitary minimal model ${\cal M}_{\nu}$, whereas the critical $RSOS(q_{1};p)$
model corresponds to the generalized $SU(2)$ coset model  ${\cal M}(q_{1},\nu
-q_{1} -2) \equiv {SU(2)_{q_{1}} \otimes SU(2)_{\nu -q_{1} -2} \over SU(2)_{\nu-2}}$ \cite{BR}. It has been also recently shown \cite{doikou} that the effective
conformal field theory for the generalized critical $RSOS(q_{1}, q_{2}; p)$ model
($q_{2} > q_{1}$) consists
of two copies of $SU(2)$ coset models, namely ${\cal M}(q_{1},\nu
-q_{1} -2) \otimes {\cal M}(q_{1},\delta q)$.
Our aim is to study the boundary behavior of the generalized $RSOS(q_{1}, q_{2};p)$
model, with ``Kondo type'' boundaries. In particular, the corresponding boundary free
energy and the $g$--function will be derived, and their relevance to boundary
flows of conformal field theories \cite{lesage, ahnrim} will be discussed.

To describe the model, an orthogonal lattice of $2N+1$
horizontal and $M$ vertical sites is considered. The Boltzmann
weights  associated with every site are defined as
\begin{equation}
w(l_{i},l_{j},l_{m},l_{n}\vert \lambda)\equiv \left(
           \begin{array}{cc}
             l_{n}   &l_{m}         \\
              l_{i}   & l_{j}      \\
                                                         \end{array}\right)\,.\non\\
\label{boltzmann}
\end{equation}
With every face
$i$ of the lattice an integer $l_{i}$ is associated, and every
pair of adjacent integers satisfies the following restriction
conditions \cite{baxter}, \cite{abf} \be 0\leq l_{i+1} - l_{i} +P \leq 2P, (a) \non\\ P \leq l_{i+1} + l_{i} \leq 2\nu -P, (b) \label{condition} \ee where
$P=q_{1}$ for $i$ odd, $P=q_{2}$ for $i$ even (let $q_{2} >
q_{1}$), for $i=1,\dots, 2N$ and $P=q$ for $i=2N+1$ for the horizontal pairs, while $P=p$
for the vertical
pairs (array type II \cite{rs}).

 The fused Boltzmann weights
have been derived by Date $\it etal$ in \cite{djmo} and they are
given by
\be
w^{q_{i},1}(a_{1}, a_{q_{i}+1},b_{q_{i}+1},b_{1}\vert \lambda)
=\sum_{a_{2} \dots a_{q_{i}}} \prod_{k=1}^{q_{i}}w^{1,1}(a_{k},
a_{k+1},b_{k+1},b_{k}\vert \lambda +i(k-q_{i}))
\ee
where $b_{2} \dots b_{q_{i}}$ are arbitrary numbers satisfying
$\vert b_{i} -b_{i+1} \vert =1$. $w^{1,1}$ are the Boltzmann weights for the $SOS(1,1)$
model \cite{baxter}, they are
non vanishing as long as the condition (\ref{condition}(a)) is
satisfied, and for $P=1$ they are given by the following expressions \be
w(l,l\pm 1,l,l\mp 1 \vert \lambda)&=& h(i-\lambda)  \non\\
w(l\pm 1,l,l\mp 1,l \vert \lambda)&=& -h(\lambda){h_{l+1} \over h_{l}}\non\\
w(l\pm 1,l,l\pm 1,l \vert \lambda)&=& h(w_{l}\pm \lambda){h_{1}
\over h_{l}} \label{boltzmann2} \ee where, \begin{equation}
h(\lambda) =\rho \Theta(\lambda)H(\lambda)
\end{equation} $H(\lambda)$ and $\Theta(\lambda)$ are Jacobi theta functions and,
\begin{equation} h_{l} =h(w_{l}), ~~~w_{l} =w_{0}+il. \end{equation}
 We are interested in the critical case where $h(\lambda)$ becomes a simple
 hyperbolic function i.e., \be
h(\lambda) = {\sinh \mu \lambda \over \sin \mu}, \label{h} \ee
 $w_{0}$,
$\rho$ and $\mu$ are arbitrary constants.
 Furthermore,
\be w^{q_{i}, p}(a_{1},
b_{1},b_{q+1},a_{q+1})=\prod_{k=0}^{p-2}\prod_{j=0}^{q_{i}-1} \Big
( h( i(k-j)+\lambda) \Big )^{-1} \non\\ \sum_{a_{2} \ldots a_{q}}
\prod_{k=1}^{p}w^{q_{i},1}(a_{k}, b_{k},b_{k+1},a_{k+1}\vert
\lambda+i (k-1)), \ee again $b_{2} \dots b_{q_{i}}$ are arbitrary
numbers satisfying $\vert b_{i} -b_{i+1} \vert =1$, and the pairs
$a_{1}$, $a_{q+1}$ and $b_{1}$,  $b_{q+1}$ satisfy
(\ref{condition}), for $P=q$. The fused weights satisfy the
Yang--Baxter equation in the following form \be \sum_{g}
w^{pq}(a,b,g,f\vert \lambda)w^{ps}(f,g,d,e\vert
\lambda+\mu)w^{qs}(g,b,c,d\vert \mu) \non\\ = \sum_{g}
w^{qs}(f,a,g,e\vert \mu)w^{ps}(a,b,c,g\vert \lambda+
\mu)w^{pq}(g,c,d,e\vert \lambda). \label{yb2} \ee Here we only
need the explicit expressions for $w^{q_{i},1}$ which are
 \be
w^{q_{i},1}(l+1,l'+1,l',l \vert
\lambda))=h_{q_{i}-1}^{q_{i}-1}(-\lambda)h_{a}
{h(ib-\lambda)\over h_{l}} \non\\
w^{q_{i},1}(l+ 1,l'-1,l',l \vert \lambda))=h_{q_{i}-1}^{q_{i}-1}(-\lambda)h_{b}{h(\lambda+ia)
\over h_{l}}\non\\
w^{q_{i},1}(l-1,l'+1,l',l \vert
\lambda))=h_{q_{i}-1}^{q_{i}-1}(-\lambda)h_{c}{h(id-\lambda) \over h_{l}} \non\\
w^{q_{i},1}(l- 1,l'-1,l',l \vert
\lambda))=h_{q_{i}-1}^{q_{i}-1}(-\lambda)h_{d}{h(ic-\lambda) \over h_{l}}
\label{boltzmann3} \ee where \be a={l+l'-q_{i} \over
2},~~~b={l'-l+q_{i} \over 2}, ~~~c={l-l'+q_{i} \over 2},
~~~d={l+l'+q_{i} \over 2}, \ee
and
\be
h_{k}^{q}(\lambda) = \prod_{j=0}^{q-1} h\Big(\lambda+i(k-j)\Big). \label{h1}\ee
 It is obvious that $w^{q_{i},1}(a,b,c,d\vert \lambda)$ are periodic functions,
 because they involve only simple hyperbolic functions (\ref{boltzmann3}), (\ref{h})
$h(\lambda+i\nu) =-h(\lambda)$, $\nu = {\pi \over \mu}$), i.e.
 \be
w^{q_{i},1}(a,b,c,d\vert \lambda +i\nu) = (-)^{q_{i}}w^{q_{i},1}(a,b,c,d\vert \lambda).
\label{period} \ee
Now we can define the transfer matrix of the $RSOS(q_{1},q_{2};p)$
model \be T^{q_{1},q_{2}, q;p\{b_{1}\ldots b_{2N+1} \}}_{\{a_{1} \ldots
a_{2N+1}\}} &=& \prod_{j=1}^{2N-1}w^{q_{1},p}(a_{j},a_{j+1},
b_{j+1},b_{j}\vert \lambda)w^{q_{2},p}(a_{j+1},a_{j+2},
b_{j+2},b_{j+1}\vert \lambda)\non\\ & &w^{q,p}(a_{2N+1},a_{2N+2},
b_{2N+2},b_{2N+1}\vert \lambda)\ee where we impose periodic boundary
conditions, i.e. $a_{2N+2} =a_{1}$ and $b_{2N+2} =b_{1}$. By finding the
eigenvalues of the transfer matrix we end up with the Bethe ansatz equations of the model,
(see also \cite{BR, doikou})
\begin{equation}
\omega^{-2}e_{q}(\lambda_{\alpha}-\Theta)e_{q_{1}}(\lambda_{\alpha})^{N}
e_{q_{2}}(\lambda_{\alpha})^{N}=
-\prod_{\beta=1}^{M} e_2(\lambda_{\alpha}-\lambda_{\beta})
\label{BAER}
\end{equation}
  It is important to mention that the eigenstates
of the model are states which satisfy (see also \cite{BR, rs, doikou})
\begin{equation}
M={1\over 4}(q_{1} + q_{2})L +{q\over 2}\\. \label{spinr}
\end{equation}
To formulate the thermodynamic Bethe ansatz for the $RSOS(q_{1}, q_{2};p)$ model we consider all the strings (\ref{STR}) with length  $ n=1, \ldots ,\nu -2$, \cite{BR, doikou}. The densities of the corresponding holes and pseudo--particles are derived from the
Bethe ansatz equations (\ref{BAER}), after we insert all the allowed strings, and they satisfy  \be  \tilde
\rho_{n}(\lambda)= {1\over 2}(Z_{nq_{1}}^{(\nu)}(\lambda
)+Z_{nq_{2}}^{(\nu)}(\lambda)) + {1\over L}
Z_{nq}^{(\nu)}(\lambda-\Theta)  - \sum_{m=1}^{\nu
-2}A_{nm}^{(\nu)}
* \rho_{m}(\lambda).
\label{densitiesr}, \ee where $L=2N$. From the constraint (\ref{spinr}) it follows that
$\tilde \rho_{\nu-2} =0$, then the density of the $\nu -2$ string can be
written
in terms of the remaining densities
 \be
\rho_{\nu -2}(\lambda) = \rho^{0}(\lambda)-\sum_{m=1}^{\nu-3}
a_{\nu -2 -m}^{(\nu -2)}*\rho_{m}(\lambda) \label{ro} \ee where
$a_{n}^{\nu-2}$ is given in the appendix (\ref{ro0}) with $\nu \to \nu -2$, and \be \hat \rho^{0}(\omega) =
 {\sinh (q_{1}{ \omega \over 2}) + \sinh
(q_{2}{ \omega \over 2}) \over 4 \cosh({\omega \over 2})
\sinh((\nu -2){\omega \over 2})} + {1\over L}{\sinh (q{ \omega \over 2}) \over 2
\cosh({\omega \over 2})
\sinh((\nu -2){\omega \over 2})}.
\ee By means of the relation (\ref{ro}) the equation
(\ref{densitiesr}) can be rewritten in the following form \be
\tilde \rho_{n}(\lambda)= {1\over 2}(Z_{nq_{1}}^{(\nu
-2)}(\lambda)+ Z_{nq_{2}}^{(\nu -2)}(\lambda))+ {1\over L}Z_{nq}^{(\nu
-2)}(\lambda-\Theta) - \sum_{m=1}^{\nu
-3}A_{nm}^{(\nu -2)}
* \rho_{m}(\lambda), \label{densities2r} \ee
$Z_{nm}^{(\nu-2)}$, $A_{nm}^{(\nu-2)}$ are given by (\ref{Z}),
(\ref{A}), with $\nu \to \nu -2$. Following the standard procedure
of minimizing the free energy of the system ($\delta f =0$, $f$ is derived by (\ref{entropy})) we obtain
the thermodynamic Bethe ansatz equations which are the same as in
the bulk \cite{doikou} \be T\ln \Big (1+\eta_{n}(\lambda)\Big ) =
-{1\over 2}(Z_{nq_{1}}^{(\nu -2)}(\lambda)+Z_{nq_{2}}^{(\nu
-2)}(\lambda)) + \sum_{m=1}^{\nu -3} A_{nm}^{(\nu -2)}* T\ln \Big
(1+\eta_{m}^{-1}(\lambda)\Big ), \label{TBAR} \ee alternatively we
can write \be \epsilon_{n}(\lambda) &=&s(\lambda)*T \ln (1+
\eta_{n+1}(\lambda))(1+ \eta_{n-1}(\lambda)) -{1\over
2}s(\lambda)(\delta_{nq_{1}} +\delta_{nq_{2}}). \label{TBA2R} \ee
 The
corresponding free energy is \be f(T)= e_{0} + f_{0} +f_{b}
+O({1\over L}) \label{contributionr} \ee where $e_{0}$ is given by
(\ref{energyim}) and $f_{0}$ is the bulk part of the free energy
given by (\ref{fe0}). We wish to compute the non--trivial boundary
part of the free energy, which is given as in the previous case
---for the alternating spin chain--- by (\ref{freeboundary1}).
Again we can evaluate differences of free energies at $T \to 0$
and $T\to \infty$. For this purpose we need the solutions of $1+\eta_{q}$
(\ref{TBA2R}), for $T \to \infty$ and $T\to 0$. For $T \to
\infty$, the main contribution in (\ref{freeboundary1}) comes for
$\lambda \to \infty$ \cite{BR, doikou} the solution is \be
(1+\eta_{n}^{\infty})^{-1} ={\sin ^{2}({\pi \over \nu}) \over \sin
^{2}({\pi (n+1) \over \nu})}, ~~n=1, \ldots, \nu -3. \label{solr1}
\ee and for $T\to 0$ the main contribution to the free energy
(\ref{freeboundary1}) comes for $\lambda \to -\infty$, and the
corresponding solution is \cite{doikou} \be (1+\eta_{n}^{0})^{-1}
&=& {\sin ^{2}({\pi \over q_{1} +2}) \over \sin ^{2}({\pi (n+1)
\over
q_{1} +2})}, ~~n=1, \ldots, q_{1}-1, ~~(1+\eta_{q_{1}}^{0}) =1 \non\\
(1+\eta_{n}^{0})^{-1}&=& {\sin ^{2}({\pi \over q_{2} - q_{1} + 2}) \over
\sin ^{2}({\pi (n-q_{1}+1) \over q_{2}-q_{1} +2})},
~~n=q_{1}+1, \ldots, q_{2}-1, ~~(1+\eta_{q_{2}}^{0}) =1 \non\\
(1+\eta_{n}^{0})^{-1}&=& {\sin ^{2}({\pi \over \nu-q_{2}}) \over
\sin ^{2}({\pi (n - q_{2} +1) \over \nu -q_{2}})}, ~~n=q_{2}+1,
\ldots, \nu-3. \label{solr2} \ee
We observe that the solution
at $T \to \infty$ does not depend on $q_{i}$, while at $T \to 0$
the solution clearly depends on the values of $q_{i}$.
Having in mind the above
solutions we are ready to derive ratios of $g$ functions, in
particular \be \ln{g_{\infty} \over g_{0}} =  {1\over
2}\ln{(1+\eta_{q}^{\infty})\over (1+\eta_{q}^{0})},
\label{resultr} \ee where $1+\eta_{q}^{\infty, 0}$ are given by
(\ref{solr1}), (\ref{solr2}).

It is worth making some remarks concerning (\ref{solr1}),
(\ref{solr2}), and (\ref{resultr}). In the special case where
$q_{1} =q_{2} =1$ we recover the results of \cite{lesage} for
boundary flows in minimal models ${\cal M}_{\nu =m+1}$, ($a =q-1$), whereas for
$q_{1}=q_{2} >1$ we recover the results of \cite{ahnrim} for
boundary flows in generalized $SU(2)$ coset models ${\cal M}(q_{1}, \nu
-q_{1}-2) \equiv {\cal M}(k,l)$. The corresponding $S$ matrix for the generalized coset
${\cal M}(k,l)$ model conjectured in \cite{Bernard} is given by
\be S = S_{RSOS}^{(k)}\otimes S_{RSOS}^{(l)}, \ee and in the limit
$l\to \infty$ it reduces to $S_{WZW_{k}}$ given in (\ref{sw}).
Again from (\ref{solr1}), (\ref{solr2}) and (\ref{resultr})
the structure ${\cal M}(q_{1},\nu
-q_{1} -2) \otimes
{\cal M}(q_{1},\delta q)$ of the effective conformal field theory is manifest
(compare with the boundary flow in $SU(2)$ coset models \cite{ahnrim}).

\section{Open Boundaries}

The main aim of this section is the investigation of the thermodynamics of the
alternating spin chain in the presence of integrable boundaries. To construct
the spin chain with boundaries in addition to the $R$ matrix
another constructing element, the $K$ matrix, is needed. The $K$
matrix is a solution of the reflection (boundary Yang--Baxter)
equation \cite{cherednik}, \be R_{12}(\lambda_{1}-\lambda_{2})\
K_{1}(\lambda_{1})\
R_{2 1}(\lambda_{1}+\lambda_{2})\ K_{2}(\lambda_{2}) \non \\
= K_{2}(\lambda_{2})\  R_{12}(\lambda_{1}+\lambda_{2})\
K_{1}(\lambda_{1})\ R_{21}(\lambda_{1}-\lambda_{2}) \,.
\label{BYBE} \ee
In what follows we are going to use Sklyanin's formalism
\cite{sklyanin} in order to construct the model with boundaries.
The corresponding transfer matrix $t(\lambda)$ for the open
alternating chain of $N$ sites and $S_{1}={q_{1}\over 2}$,
$S_{2}={q_{2}\over 2}$ spins is (see also e.g., \cite{sklyanin, VEWO}),  \be t(\lambda) = \tr_{0} K_{0}^{+}(\lambda)\
T_{0}(\lambda)\ K^{-}_{0}(\lambda)\ \hat T_{0}(\lambda)\,,
\label{transfer2} \ee where\footnote{We could have considered $N$=odd and put an
impurity at the $N$ site of the chain. Then we would have an
impurity contribution to the free energy similar to the one in
paragraph 2.} \be T_{ 0}(\lambda) = R_{0N}^{1}(\lambda) R_{0N-1
}^{2}(\lambda)\cdots R_{02}^{1}(\lambda) R_{01}^{2}(\lambda),
~~\hat T_{ 0}(\lambda) = R_{10}^{2}(\lambda) R_{20
}^{1}(\lambda)\cdots R_{N-10}^{2}(\lambda) R_{N0}^{1}(\lambda),
\label{hatmonodromy} \ee and $K^{+}(\lambda, \xi^{+}, \kappa^{+}) = K^{-}(-\lambda -i, \xi^{-}, \kappa^{-})^{t}$ where $\xi^{\pm}$, $\kappa^{\pm}$ are arbitrary boundary parameters for the left and right boundaries, and $K^{-}$ is the matrix \cite{GZ, DVGR3}
\be K^{-}(\lambda) = \left(
         \begin{array}{cc}
           \sinh \mu(\lambda +i\xi)  &\kappa \sinh \mu 2\lambda        \\
           \kappa \sinh \mu 2\lambda    &\sinh \mu(-\lambda +i\xi)  \\
                                                        \end{array}\right)\,.
                                                       \label{gz}   \ee It is interesting to point out that the
diagonal boundaries for the critical $XXZ$ spin chain
($S_{1}=S_{2} ={1\over 2}$) correspond to Dirichlet boundary
conditions for the sine--Gordon model \cite{lmss}, \cite{fsw}.
Presumably the purely anti--diagonal $K$ matrix
---which completely breaks the $U(1)$ symmetry--- should
correspond to Neumann boundary conditions for the sine--Gordon
model.

\subsection{The diagonal $K$ matrix}

We consider first the case in which the $K$ matrix is diagonal,
namely $\kappa =0$. The corresponding Bethe ansatz for the model,
are known \cite{bado} and they are given by
\begin{equation}
e_{x^{+}}(\lambda_{\alpha})^{-1}e_{x^{-}}(\lambda_{\alpha})e_{1}(\lambda_{\alpha})
g_{1}(\lambda_{\alpha})e_{q_{1}}(\lambda_{\alpha})^{N}e_{q_{2}}(\lambda_{\alpha})^{N}
= -\prod_{\beta=1}^{M} e_2(\lambda_{\alpha}-\lambda_{\beta})
e_2(\lambda_{\alpha}+\lambda_{\beta}) \label{BEO}
\end{equation}
where $x^{\pm}=2\xi ^{\pm}\pm 1$ are boundary
parameters for the left right boundaries of the chain and they are
related to some external magnetic field acting on the boundaries.
Here, for simplicity we consider $x^{\pm}$ to be integers less or
equal to $\nu -1$.
The corresponding densities of pseudo--particles and holes, in analogy to
(\ref{densitiesi}) satisfy
(again we consider as in the bulk all the strings $n =1, \ldots ,\nu-1$ and the negative parity string (\ref{STR}))
\be & & \tilde \rho_{n}(\lambda)=Z_{nq_{1}}(\lambda)+
Z_{nq_{2}}(\lambda) +{1\over L} K_{n} - \sum_{m=1}^{\nu -1}A_{nm}
* \rho_{m}(\lambda) - B_{1n} * \rho_{0}(\lambda) \non\\
 &-& (\rho_{0}(\lambda) + \tilde \rho_{0}(\lambda)) =b_{q_{1}}
+b_{q_{2}} +{1\over L} K_{0} -\sum_{m=1}^{\nu -1}B_{1m} *
\rho_{m}(\lambda) -a_{2}*\rho_{0}(\lambda) \label{densities0} \ee
where $L=N$ is the length of the chain and, \be \hat K_{n}(\omega)
&=& \hat a_{n}(\omega)+ \hat b_{n}(\omega)- \hat
Z_{nx^{+}}(\omega)+\hat Z_{nx^{-}}(\omega) -1 \non \\ \hat
K_{0}(\omega) &=& \hat a_{1}(\omega)+ \hat b_{1}(\omega)- \hat
b_{x^{+}}(\omega)+\hat b_{x^{-}}(\omega) + 1 \label{kbound} \ee
$\hat Z_{nm}$, $\hat b_{n}$, and $\hat a_{n}$ are given in the
appendix by (\ref{Z}), and (\ref{ro0}). The unit that appears in
the expressions for $\hat K_{n}$, $\hat K_{0}$ is a result of the
subtraction of a  $\delta(\lambda)$ term from the densities (see
also \cite{lmss, drtw}). This subtraction seems necessary for the
accurate derivation of the density, because in the boundary case
$\lambda$'s can take values from $0$ to $\infty$, as opposed to
the bulk case where $\lambda$'s take all the values from $-\infty$
to $\infty$.

The free energy for the model with open boundaries is given by
(\ref{entropy}) up to a ${1\over 2}$ factor in front of the
expression. The appearance of the factor ${1\over 2}$ in front of
all the integrals comes from the fact that we originally derived
the integrals from zero to infinity. After we minimize the free
energy, we obtain the same thermodynamic Bethe ansatz equations as
in (\ref{TBA}), ({\ref{TBA2}). Finally, by virtue of (\ref{TBA}),
({\ref{TBA2}) the following expression for the free energy is
obtained \be f=f_{0}+f_{b} +O({1\over L}) \ee where $f_{0}$ is the
bulk free energy given by \be f_{0} = e_{0}- {T\over
2}\int_{-\infty}^{\infty} d \lambda
s(\lambda)\ln(1+\eta_{q_{1}}(\lambda))(1+\eta_{q_{2}}(\lambda))
\ee $e_{0}$ is the bulk part of the energy of the state with the
seas of strings $q_{1}$, $q_{2}$ filled, \be e_{0}= - {1\over
4}\sum_{i, j=1}^{2}\int_{-\infty}^{\infty}d\lambda
Z_{q_{i}q_{j}}(\lambda)s(\lambda). \ee
 and the boundary contribution to the free energy is
\be f_{b} = -{T\over 2L} \sum_{n=1}^{\nu
-1}\int_{-\infty}^{\infty}d\lambda
K_{n}(\lambda)\ln(1+\eta_{n}^{-1}(\lambda)) +{T\over 2L}
\int_{-\infty}^{\infty}d\lambda
K_{0}(\lambda)\ln(1+\eta_{0}^{-1}(\lambda)). \label{freeboundary2}
\ee From the thermodynamic Bethe ansatz equations (\ref{TBA}) we
consider the convolution of $s(\lambda)$ with the equation of
(\ref{TBA}) for $n=1$ and $n=x^{\pm}$ (with a (-) sign in front of
the equation for  $n=x^{+}$), we also consider the convolution of
$s(\lambda)$ with the equation for the negative parity string,
recall also relations (\ref{ident}). Moreover, we consider the
equations (\ref{TBA2}) for all $n =1, \ldots ,\nu-1$ and the
negative parity string, with a (-) sign infront. We add all the
above equations and we end up with the following expression for
the free energy (at $ T \to 0, \infty $) \be f_{b}=e_{b} + \delta f_{b} =e_{b} +{T\over L}
\ln(1+\eta_{\nu -1}) +{T\over 2L}s* \ln(1+\eta_{x^{+}})-{T\over
2L}s* \ln(1+\eta_{x^{-}}) \ee where $e_{b}$ is the boundary energy
contribution  \be e_{b} &=& -{1\over 4L}\int_{-\infty}^{\infty}
d\lambda s(\lambda)\Big ( K_{q_{1}}(\lambda)+
K_{q_{2}}(\lambda)\Big ) \non\\ &=&  -{1\over
2L}\int_{-\infty}^{\infty} d\lambda s(\lambda) \Big ({1\over 2}
\sum_{i=1}^{2}
(a_{q_{i}}(\lambda)+b_{q_{i}}(\lambda)-Z_{q_{i}x^{+}}+Z_{q_{i}x^{-}})-
\delta(\lambda)\Big ).\ee The boundary contribution to the free
energy for each boundary then is \be \delta f_{b}^{\pm}={T\over 2
L} \ln(1+\eta_{\nu -1}) \pm {T\over 4L} \ln(1+\eta_{x^{\pm}}),
\label{bcont}\ee and the corresponding ratios of the
$g^{+},~g^{-}$ ($g=g^{+}g^{-}$) functions for the left and right
boundaries (\ref{gfunction}) are \be \ln{g_{\infty}^{\pm} \over
g_{0}^{\pm}} = - {1\over 2}\ln{(1+\eta_{\nu -1}^{\infty})\over
(1+\eta_{\nu -1}^{0})}  \mp {1\over
4}\ln{(1+\eta_{x^{\pm}}^{\infty})\over (1+\eta_{x{\pm}}^{0})}.
\label{bcont2} \ee The last term of the above equation, for
$x^{\pm}$ integer less than $\nu -1$, corresponds to the quantum
impurity contribution (up to a $-{1\over 2}$ factor), which has
been already computed explicitly for $T \to \infty ,0$,
(\ref{sol1}), (\ref{sol2}). We focus on the first term of
(\ref{bcont2}), which we compute again for $T \to \infty ,0$. It
is easy to conclude from the solution for $T\to \infty ,0$
(\ref{sol1}), (\ref{sol2}) \be - {1\over 2}\ln{(1+\eta_{\nu
-1}^{\infty})\over (1+\eta_{\nu -1}^{0})} = -{1\over 2} \ln {\nu
\over \nu -q_{2}}. \label{final} \ee The result (\ref{final}) is
also valid for the spin $S=S_{1}=S_{2}$ chain. In the isotropic
limit the above ratio (\ref{final}) becomes unit, i.e. there is no
difference in the boundary free energy at low and high
temperature. In the special case where $S_{1}=S_{2}={1\over 2}$ we
recover the result found in \cite{fsw} for the sine--Gordon model
with boundaries \footnote{In the repulsive regime, i.e. when no
bound states ``breathers'' exist, we can make the following
identification ${\nu  \over \nu -1}= \lambda +1$  or $ \beta^{2} =
8(\pi -\mu)$ \cite{fsw, doikou-nepo}.}, and also the result of
\cite{desa} for the $XXZ$ spin chain with open boundaries at zero
external magnetic field. Note that for $x^{\pm} \to \infty$ there
is no boundary
parameter dependence. 

\subsection{The non--diagonal $K$ matrix}

We consider, for the first time in a spin chain model, the more
general case of the open alternating spin chain with non--diagonal
boundaries. The corresponding Bethe ansatz for the $XXZ$ model at
roots of unity have been recently derived in \cite{nepo1}. The
novelty of the method described in \cite{nepo1} is basically that
it does not rely on the existence of a reference state
``pseudo--vacuum''. It was shown then in \cite{nepo1}, that in the
case where $\mu={\pi \over p+1}$ ($\nu =p+1$) the problem of
finding the eigenvalues of the transfer matrix (\ref{transfer2})
reduces to a set of functional relations which can be written in
the following compact form, (see also \cite{BR, doikou, nepo1})
\be detM[\Lambda^{q_{1},q_{2}}(\lambda)]=0 \ee where
\begin{equation}
M[\Lambda^{q_{1},q_{2}}(\lambda)]= \left(
        \begin{array}{cccccccc}
 \Lambda_{0}^{q_{1},q_{2}} &-\tilde f_{-1}^{q_{1},q_{2}} &0 &0 &\ldots &0 &0  &- f_{0}^{q_{1},q_{2}}    \\
 -f_{1}^{q_{1},q_{2}} &\Lambda_{1}^{q_{1},q_{2}} &-\tilde f_{0}^{q_{1},q_{2}} &0 &\ldots &0 &0 &0  \\
 0 &-f_{2}^{q_{1},q_{2}} &\Lambda_{2}^{q_{1},q_{2}} &-\tilde f_{1}^{q_{1},q_{2}} &\ldots &0 &0 &0  \\
 \vdots &\vdots   &\vdots   &\vdots    &\ldots  &\vdots   &\vdots   &\vdots  \\
 0 &0 &0 &0 &\ldots &-f_{p-1}^{q_{1},q_{2}} &\Lambda_{p-1}^{q_{1},q_{2}} &-\tilde f_{p-2}^{q_{1},q_{2}}  \\
-\tilde f_{p-1}^{q_{1},q_{2}} &0 &0 &0 &\ldots &0 &-f_{p}^{q_{1},q_{2}}&\Lambda_{p}^{q_{1},q_{2}} \\                       \end{array}\right)\,,\non\\
\label{matrix2}
\end{equation}
where
\be
f^{q_{1},q_{2}}(\lambda) & =& -\sinh^{N}\mu (\lambda+iS_{1} +{i\over 2}) \sinh^{N} \mu (\lambda+iS_{2} +{i\over 2}){\sinh \mu (2\lambda+2i) \over \sinh \mu (2\lambda+i)}\non\\ & &\Big (\sinh \mu (\lambda+i\xi)\sinh \mu (\lambda-i\xi)+\kappa^{2} \sinh^{2} 2\mu \lambda \Big)
\label{f1} \ee
and
\be
\tilde f^{q_{1},q_{2}}(\lambda) = f^{q_{1},q_{2}}(-\lambda -2i),~~f_{k}^{q_{1},q_{2}}(\lambda)=f(\lambda+ik) \label{f2} \ee
Let now $(Q_{0}(\lambda), \ldots ,Q_{p}(\lambda))$ be
the null vector of the matrix (\ref{matrix2}) with $Q_{k}(\lambda)=Q(\lambda+ik)$
and
\be
Q(\lambda) = \prod_{j=1}^{M}
\sinh \mu(\lambda-\lambda_{j})\sinh \mu(\lambda+\lambda_{j}+i)
 \ee
then the eigenvalues are given by the following expression
\be \Lambda^{q_{1},q_{2}}(\lambda) =
f_{0}^{q_{1},q_{2}}(-\lambda-i) {Q(\lambda
+i) \over Q(\lambda)} +
f_{0}^{q_{1},q_{2}}(\lambda){Q(\lambda-i) \over Q(\lambda)}.
\label{eigen} \ee
Finally, from the analyticity of the eigenvalues we obtain the Bethe ansatz equations
\be &&{\sinh \mu (\lambda_{\alpha}
+ {i \over 2}(2\xi-1))\sinh \mu (\lambda_{\alpha} - {i \over
2}(2\xi+1)) + \kappa^{2}\sinh^{2} \mu (2 \lambda_{\alpha}-i) \over
\sinh \mu (\lambda_{\alpha} - {i \over 2}(2\xi-1))\sinh \mu
(\lambda_{\alpha} + {i \over
2}(2\xi+1)) + \kappa^{2} \sinh^{2} \mu (2 \lambda_{\alpha} +i)} \non \\
&&
g_{1}(\lambda_{\alpha})e_{1}(\lambda_{\alpha})e_{q_{1}}(\lambda_{\alpha})^{N}
e_{q_{2}}(\lambda_{\alpha})^{N}= -\prod_{\beta=1}^{M}
e_2(\lambda_{\alpha}-\lambda_{\beta})
e_2(\lambda_{\alpha}+\lambda_{\beta}). \label{be2} \ee Note that
the boundary parameters at the left and right boundaries have been
considered to be the same. In fact, the boundary parameters have
to be tuned properly (i.e. $\xi^{-} =\xi^{+}$ and $\kappa^{-} =
\kappa^{+}$) so that the method described in \cite{nepo1} can be
applied. Note also that in this case the Bethe ansatz states have
to satisfy the following constraint, \be M=(S_{1}+S_{2}){N\over
2}-{1\over 2}. \label{restr} \ee The above constraint follows from
the asymptotic behavior of the transfer matrix for $\lambda \to
\infty$, in particular from (\ref{transfer2}) it can be deduced
\be t(\lambda \to \infty) \sim - 2 e^{(2\mu N +4\mu)\lambda +2i\mu
+i\mu N} \label{asympt} \ee by comparing the later equation with
the eigenvalues (\ref{eigen}) we obtain the constraint
(\ref{restr}).

Let us first consider the case with purely anti--diagonal $K$
matrices, namely $\kappa$ becomes very big. Then it is obvious
that the above Bethe ansatz equations (\ref{be2}) take the form
\be
g_{1}(\lambda_{\alpha})^{-2}e_{1}(\lambda_{\alpha})^{-2}g_{1}(\lambda_{\alpha})e_{1}(\lambda_{\alpha})e_{q_{1}}(\lambda_{\alpha})^{N}
e_{q_{2}}(\lambda_{\alpha})^{N} = -\prod_{\beta=1}^{M}
e_2(\lambda_{\alpha}-\lambda_{\beta})
e_2(\lambda_{\alpha}+\lambda_{\beta}). \label{2} \ee  Relation
(\ref{restr}) does not impose in this case any extra restriction
on the densities as opposed to the case of the $RSOS$ model for
which a similar relation (\ref{spinr}) holds true, and it imposes
restrictions on the hole densities (i.e. $\tilde \rho_{\nu-2}
=0$). Therefore, the thermodynamic Bethe ansatz equations are the
same as in the bulk case (\ref{TBA}), and by following exactly the
same procedure as in the diagonal case, but now with \be \hat
K_{n}(\omega) = - \hat b_{n}(\omega)- \hat a_{n}(\omega) -1,
~~\hat K_{0}(\omega) = - \hat b_{1}(\omega)-\hat a_{1}(\omega) + 1
\label{kbound2} \ee we find that the non--trivial boundary part of
the free energy becomes (at $ T \to 0, \infty $)  \be \delta f_{b}^{\pm}={T\over 4 L}
\ln(1+\eta_{1}).\label{bcont1} \ee Moreover the boundary
contribution to the ground state energy is given by \be e_{b} =
-{1\over 2L}\int_{-\infty}^{\infty} d\lambda s(\lambda) \Big
({1\over 2} \sum_{i=1}^{2}
(-a_{q_{i}}(\lambda)-b_{q_{i}}(\lambda)) -\delta(\lambda)\Big
).\ee It is worth noticing that the state with the seas of strings
with length $q_{1}$, $q_{2}$ filled, is an allowed state for the
case with anti--diagonal boundaries, because the constraint
(\ref{restr}) is satisfied. The ratios of the $g^{+},~g^{-}$
functions for the left and right boundary (\ref{gfunction}) are
given by \be \ln{g_{\infty}^{\pm} \over g_{0}^{\pm}} =   - {1\over
4}\ln {(1+\eta_{1}^{\infty})\over (1+\eta_{1}^{0})}
\label{bcont22} \ee where the function
$(1+\eta_{1}^{0,\infty})^{-1}$ is given, for $T \to \infty$ and
$T\to 0$, by (\ref{sol1}), (\ref{sol2}). Notice that for $q_{i}
=1$ (at $T\to 0 $) $\delta f_{b}^{\pm} \propto T^{2}$ and this is
---as in the presence of an impurity at the end of the chain---
the completely screened case.

We are not going to treat the general case in full detail here,
since this is mostly a problem of mathematical manipulations, he
hope though to report on this in detail elsewhere. Let us however
present the general concept of such computation for the special
simple case where $\xi =0$, the so called ``free boundary
conditions'' \cite{GZ}, then the Bethe ansatz equations become \be
e_{2\zeta+1}(\lambda_{\alpha})^{-1}e_{2\zeta-1}(\lambda_{\alpha})
e_{1}^{-2}(\lambda_{\alpha})e_{1}(\lambda_{\alpha})g_{1}(\lambda_{\alpha})e_{q_{1}}(\lambda_{\alpha})^{N}e_{q_{2}}(\lambda_{\alpha})^{N}
= -\prod_{\beta=1}^{M} e_2(\lambda_{\alpha}-\lambda_{\beta})
e_2(\lambda_{\alpha}+\lambda_{\beta}). \label{BEO2} \ee It is
evident that the later Bethe ansatz equations have a similar
structure with (\ref{BEO}) ---up to the $e_{1}^{-2}$ term in the
LHS of (\ref{BEO2})---  with $\xi^{+} =\xi^{-}=\zeta$ and
$\cosh^{2} \mu i\zeta =-{1\over 4 \kappa^{2}}$ \footnote{Following
\cite{GZ} we have to chose for the ``free boundary conditions''
$\zeta = {1 \over 2} {\nu^{2} \over \nu -1}$, which is not an
integer}. The boundary part of the free energy is given again by
(\ref{freeboundary2}) and the quantities $K_{n}$ and $K_{0}$ are
given by, \be \hat K_{n}(\omega) &=& -\hat a_{n}(\omega)+ \hat
b_{n}(\omega)- \hat Z_{nx^{+}}(\omega)+\hat Z_{nx^{-}}(\omega) -1
\non \\ \hat K_{0}(\omega) &=& \hat a_{1}(\omega)- \hat
b_{1}(\omega)- \hat b_{x^{+}}(\omega)+\hat b_{x^{-}}(\omega) + 1
\label{kbound3} \ee with $x^{\pm} =2 \zeta \pm 1$. It remains to
treat expression (\ref{freeboundary2}) at high and low
temperature, but this is relatively easy to do because the
solutions for $\eta_{n}$, $\eta_{0}$ are known at $T \to \infty$
and $T \to 0$ (\ref{sol1}), (\ref{sol2}). Basically we have to
focus on the part of expression (\ref{freeboundary2}) that
involves the terms $\pm Z_{nx^{\mp}}$, $\pm b_{x^{\mp}}$ for
$x^{\pm}$ non--integer. The remaining $x^{\pm}$ independent part
of the boundary free energy (\ref{freeboundary2}), denoted as
$f_{b}^{(1)}$, can be easily computed by repeating the steps of
the previous section, and it is given by (at $ T \to 0, \infty $)  \be f_{b}^{(1)} =
-{1\over 2L}\int_{-\infty}^{\infty} d\lambda s(\lambda) \Big
({1\over 2} \sum_{i=1}^{2}
(-a_{q_{i}}(\lambda)+b_{q_{i}}(\lambda))- \delta(\lambda)\Big ) +
{T\over L} \ln(1+\eta_{\nu -1})+ {T\over 2L} \ln(1+\eta_{1}).
\label{general} \ee The $x^{\pm}$ dependent part of the boundary
free energy, denoted as $f_{b}^{(2)}$ (with terms that involve
$\pm Z_{nx^{\mp}}$, $\pm b_{x^{\mp}}$), which needs to be computed
explicitly for $x^{\pm}$ non--integers, has the form \be
f_{b}^{(2)} &=& -{T\over 2L} \sum_{n=1}^{\nu
-1}\int_{-\infty}^{\infty}d\lambda (Z_{nx^{-}}(\lambda)
-Z_{nx^{+}}(\lambda))\ln(1+\eta_{n}^{-1}(\lambda)) \non\\ &+&
{T\over 2L} \int_{-\infty}^{\infty}d\lambda
(b_{x^{-}}(\lambda)-b_{x^{+}}(\lambda))
\ln(1+\eta_{0}^{-1}(\lambda)). \label{freeboundary3} \ee Again
once we make the shift $\lambda \to \lambda -{1\over \pi}\ln T $
we can compute $f_{b}^{(2)}$ analytically, at $T\to 0, \infty$
having in mind the expressions for $Z_{nm}$, $b_{n}$ (\ref{a1}),
(\ref{a2}), and $\eta_{n}$, $\eta_{0}$ at high and low
temperatures (\ref{sol1}), (\ref{sol2}). More specifically,
expression (\ref{freeboundary3}) can be written as $f_{b}^{(2)} =
f_{b}^{(2)}(x^{-}) +f_{b}^{(2)}(x^{+})$  where at $T\to 0,
\infty$, \be f_{b}^{(2)}(x^{\pm}) =\pm {T\over 2L}
\sum_{n=1}^{\nu -1} \hat
Z_{nx^{\pm}}(0)\ln(1+\eta_{n}^{-1}) \mp {T\over 2L}
\hat b_{x^{\pm}}(0) \ln(1+\eta_{0}^{-1}).
\label{freeboundary4} \ee Naturally, the case is similar when
diagonal boundaries with $x^{\pm} =2\xi \pm 1$ non--integer, are
considered, i.e. the $x^{\pm}$ dependent part of the free energy
(\ref{freeboundary2}) is given again by (\ref{freeboundary4}).
Eventually the computation of the expression (\ref{freeboundary4})
reduces to a simpler problem, which is the derivation of the exact
Fourier transforms of $Z_{nx}$ and $b_{x}$ when $x$ is
non--integer. In particular, expression (\ref{ro0}) for $\hat
b_{x}$ is valid as long as $x < \nu$, while expression (\ref{Z})
for $\hat Z_{nx}$ is also valid as long as $x>n$.

Let us point out that the results of this section are novel not only for the fused model
under study, but also for the case $q_{1} =q_{2}=1$, which corresponds to the $XXZ$ model (sine--Gordon model).

\section{Discussion}

In this investigation we focused on the  ``boundary properties''
of the alternating spin chain and the $RSOS(q_{1},q_{2};p)$ model,
therefore we considered both models with certain integrable
``boundaries''. A quantum impurity was added at the last site of
the chain (``Kondo type'' boundary see also \cite{ftw, ande}), and
the immediate result was a non--trivial contribution to the ground
state energy of the system, as well as in the free energy. We were
able to explicitly evaluate the non--trivial contribution to the
boundary free energy at low and high temperature. A similar
investigation was realized for the $RSOS(q_{1},q_{2};p)$ model
(see also \cite{doikou}) and the obtained results, for special
values of $q_{1}$, $q_{2}$, compared with the known boundary flows
in unitary minimal models \cite{lesage} and generalized $SU(2)$
coset theories \cite{ahnrim}.  For general values of $q_{1}$,
$q_{2}$ our results are rather novel for both the alternating spin
chain and the $RSOS(q_{1},q_{2};p)$ model.

Furthermore, the alternating chain with diagonal and non--diagonal
boundaries was investigated, and again the presence of the
boundaries resulted in a non--trivial contribution to the free
energy. This contribution gave rise to the $g$ function (``ground
state degeneracy'') along the lines described in \cite{af}. We
were able to derive ratios of the $g$ function for the left and
right boundaries, at low and high temperatures. In the diagonal
case for $S_{1}=S_{2}={1\over 2}$ the boundary parameter
independent part of the ratio of the $g$ functions coincides with
the one found in \cite{fsw} and \cite{desa} for zero external
magnetic field. We believe the results of this section are also
novel, in particular for the anti--diagonal case our results are
rather new even for $q_{1} =q_{2} =1$.

For completeness the thermodynamics of the $RSOS(q_{1},q_{2};p)$
model with open boundaries \cite{rsosb1, rsosb2, rsosb3} should be also investigated. It
would be also of great interest to extend the thermodynamic
analysis for spin chains and $RSOS$ models related to higher
rank algebras. Furthermore, the derivation of the
Bethe ansatz equations for non--diagonal boundaries \cite{nepo1}, is an essential step towards
the investigation of exact non--diagonal boundary $S$ matrices in the spin chain framework.
Let us finally note that a new non--trivial ``dynamical'' solution of the reflection equation has
been recently found \cite{bako}, and it has been realized in the context of the sine--Gordon model.
It is a challenging problem to apply this ``dynamical'' solution to the $XXZ$ spin chain at roots of
unity, and investigate the thermodynamics and the scattering process in this framework.  We hope to
address these questions soon in a future
work \cite{prep}.

\section{Acknowledgments}
I am grateful to F. Ravanini for helpful discussions.
This work was supported by the TMR Network
``EUCLID'';
``Integrable models and applications: from strings to condensed matter'', contract number HPRN--CT--2002--00325.

\appendix

\section{Appendix}

We give the explicit expressions of the functions that appear in the Bethe asnatz equations once we apply the string hypothesis, namely
\begin{eqnarray}
X_{nm}(\lambda)&=& e_{|n-m+1|}(\lambda) e_{|n-m+3|}(\lambda)\ldots
e_{(n+m-3)}(\lambda)e_{(n+m-1)}(\lambda)
\nonumber\\
E_{nm}(\lambda)&=& e_{|n-m|}(\lambda)e_{|n-m+2|}^{2}(\lambda)
\ldots e_{(n+m-2)}^{2}(\lambda)e_{(n+m)}(\lambda) \nonumber\\
G_{nm}(\lambda)&=& g_{|n-m|}(\lambda)g_{|n-m+2|}^{2}(\lambda)
\ldots g_{(n+m-2)}^{2}(\lambda)g_{(n+m)}(\lambda).
\label{ap1}
\end{eqnarray}
Moreover, \be a_{n}(\lambda) ={i\over 2\pi} {d\over d\lambda}\ln
e_{n}(\lambda), ~~ b_{n}(\lambda) ={i\over 2\pi} {d\over
d\lambda}\ln g_{n}(\lambda), \label{a1} \ee \be
(Z_{nm}(\lambda),A_{nm}(\lambda), B_{nm}(\lambda)) ={i\over 2\pi}
{d\over d\lambda} \ln (X_{nm}(\lambda),
E_{nm}(\lambda),G_{nm}(\lambda)).  \label{a2}  \ee We finally give
the following useful Fourier transforms \be \hat a_{n}(\omega) =
{\sinh \Big ((\nu -n){\omega \over 2}\Big) \over \sinh\Big ({\nu
\omega \over 2}\Big)}, ~~ n<2\nu, ~~\hat b_{n}(\omega)= -{\sinh
\Big ({n\omega \over 2}\Big ) \over \sinh \Big ({\nu \omega \over
2}\Big )}, ~~ n<\nu ,  \label{ro0} \ee
\begin{equation}
\hat Z_{nm}(\omega)= { \sinh \Bigl (( \nu - \max(n,m)){\omega
\over 2} \Bigr ) \sinh \Bigl ((\min(n,m)){\omega \over 2} \Bigr )
\over \sinh ({\nu \omega \over 2}) \sinh({\omega \over 2})}\\
\label{Z}
\end{equation}
\begin{equation}
\hat A_{nm}(\omega)={2 \coth ({\omega \over 2} )\sinh \Bigl (( \nu
- \max(n,m)){\omega \over 2}\Bigr) \sinh \Bigl ((\min(n,m)){\omega
\over 2}\Bigr ) \over \sinh ({\nu \omega \over 2})} \label{A}
\end{equation}
\begin{equation}
\hat B_{nm}(\omega)=-{2\coth ({\omega \over 2})\sinh\Bigl
({n\omega \over 2}\Bigr ) \sinh \Bigl ({m\omega \over 2}\Bigr)
\over \sinh ({\nu \omega \over 2})}. \label{b}
\end{equation}


\begin{thebibliography}{99}





\bibitem{cardy}
J.L. Cardy, Nucl. Phys. {\bf B234} (1989) 581.

\bibitem{af}
I. Affleck and A.W.W. Ludvig, Phys. Rev. Lett. {\bf 67} (1991) 161; I. Affleck, M. Oshikawa and H. Saleur, {\it cond--matt/9804117} (1998).

\bibitem{cherednik}
I.V. Cherednik, Theor. Math. Phys. {\bf 61} (1984) 977.

\bibitem{sklyanin}
E.K. Sklyanin, J. Phys. {\bf A21} (1988) 2375; P.P. Kulish and
E.K. Sklyanin, J. Phys. {\bf A24} (1991) L435.

\bibitem{fk}  A. Fring and  L. K\"{o}berle, Nucl. Phys. {\bf B421}
(1994) 159; Nucl. Phys. {\bf B419} (1994) 647.

\bibitem{GZ}
S. Ghoshal and A. B. Zamolodchikov, Int. J. Mod. Phys. {\bf A9}
(1994) 3841; {\bf A9} (1994) 4353.

\bibitem{DVGR3}
H.J. de Vega and A. Gonz\'alez-Ruiz, J. Phys. {\bf
A26} (1993) L519


\bibitem{lut} C.L. Kane and M.P.A. Fisher, Phys. Rev. Lett. {\bf 68} (1992) 1220; E. Sorensen, S. Eggert and I. Affleck, J. Phys. {\bf A26}
(1993) 6757.

\bibitem{lesage}  F. Lesage, H. Saleur and  P. Simonetti, Phys.Lett. {\bf B427} (1998) 85.

\bibitem{ahnrim} C. Ahn and C. Rim, J. Phys. {\bf A32} (1999) 2509.


\bibitem{ftw} V.M. Filyov, A.M. Tsvelik amd P.B. Wiegmann, Phys.
Lett. {\bf 81A} (1981) 175; A.M. Tsvelick and P.B. Wiegmann, Adv.
in Phys. {\bf 32} (1983) 453.

\bibitem{ande} N. Andrei and C. Destri, Phys. Rev. Lett. {\bf 52} (1984) 364.

\bibitem{wang}
Y. Wang, Phys. Rev. {\bf B60} (1999) 9236; Y. Wang and P. Schlottmann, Phys. Rev. {\bf B62} (2000) 3845; A. P. Tonel, A. Foerster, X.--W. Guan and J. Links, {\it cond--mat/0112115}.

\bibitem{yy}
C.N. Yang and C.P. Yang, Phys. Rev. {\bf 150} (1966) 327; J. Math.
Phys. {\bf 10} (1969) 1115.

\bibitem{y2} C.P. Yang, Phys. Rev. {\bf A2} (1970) 154.

\bibitem{g} M. Gaudin, Phys. Rev. Lett. {\bf 26} (1971) 1301.

\bibitem{t1} M. Takahashi, Prog. Theor. Phys. {\bf 46} (1971) 401; M. Takahashi and M. Suzuki, Prog. Theor. Phys. {\bf 48} (1972)
2187; M. Takahashi, {\it Thermodynamics of One--Dimensional Solvable
Models} (Cambridge University   Press, 1999).

\bibitem{jm} J.D. Johnson and B.M. McCoy, Phys. Rev. {\bf A6}
(1972) 1613.

\bibitem{BT}
H. Babujian, Nucl. Phys. {\bf B215} (1983) 317;
H. Babujian and A. Tsvelik, Nucl. Phys. {\bf B265} (1986) 24.

\bibitem{mn1}
L. Mezincescu and R.I. Nepomechie, UMTG--170 (1992).

\bibitem{z2}
Al.B. Zamolodchikov, Nucl. Phys. {\bf B342} (1990) 695; Al.B.
Zamolodchikov, Phys. Lett. {\bf B253} (1991) 391.

\bibitem{z1}
A.B. Zamolodchikov,  Nucl. Phys. {\bf B358} (1991) 497, 524; Nucl. Phys. {\bf B366} (1991) 122.

\bibitem{km}
T.R. Klassen and E. Melzer, Nucl. Phys. {\bf B338} (1990) 485.

\bibitem{dd} C. Destri, H.J. de Vega, Nucl. Phys. {\bf B438} (1995).

\bibitem{lmss} A. LeClair, G. Mussardo, H. Saleur and S. Skorik ,
Nucl. Phys. {\bf B453} (1995) 581.

\bibitem{fsw} P. Fendley, H. Saleur and  N.P. Warner, Nucl. Phys. {\bf B430} (1994) 577.

\bibitem{drtw} P. Dorey, I. Runkel, R. Tateo and G. Watts,
 Nucl. Phys. {\bf B578} (2000) 85.

\bibitem{alcaraz}
F.C. Alcaraz, M.N. Barber, M.T. Batchelor, R.J. Baxter and G.R.W.
Quispel, J. Phys. {\it A20} (1987) 6397.

\bibitem{desa} P. de Sa and A. Tsvelik, {\it cond--matt/9503031}.

\bibitem{pol} A.M. Polyakov, J.E.T.P. Lett. {\bf 12} (1970) 381.

\bibitem{bpz}
A.A. Belavin, A.M. Polyakov and A.B. Zamolodchikov, J. Stat. Phys.
{\bf 34} (1984) 763; Nucl. Phys. {\bf B241} (1984) 333.

\bibitem{bc}
H.W.J. Bl\"{o}te, J.L. Cardy, M.P. Nightingale, Phys. Rev. Lett.
{\bf 56} (1986) 742; J.L. Cardy, Nucl. Phys. {\bf B270} (1986)
186.

\bibitem{af1}
I. Affleck, Phys. Rev. Lett. {\bf 56} (1986) 746.


\bibitem{VEWO} H.J. de Vega and F. Woyanorovich, J. Phys. {\bf A25} (1992) 4499.

\bibitem{doikou} A. Doikou, J. Phys. {\bf A36} (2003) 329.

\bibitem{baxter}
R.J. Baxter, Ann.
Phys. {\bf 70} (1972) 193; J. Stat. Phys. {\bf 8} (1973) 25; {\it
Exactly Solved Models in Statistical Mechanics} (Academic Press,
1982)

\bibitem{korepin}
V.E. Korepin, Theor. Math. Phys. {\bf 76} (1980) 165; V.E.
Korepin, G. Izergin and N.M. Bogoliubov, {\it Quantum Inverse
Scattering Method, Correlation Functions and Algebraic Bethe
Ansatz} (Cambridge University Press, 1993).

\bibitem{KR} A. Kirillov and
N.Yu Reshetikhin, J. Sov. Math {\bf 35} (1986) 2621; A. Kirillov
and N.Yu Reshetikhin, J. Phys. {\bf A20} (1987) 1565.

\bibitem{FT2}
L.D. Faddeev and L.A. Takhtajan, Russ. Math. Surv. {\bf 34}, 11
(1979); L.D. Faddeev and L.A. Takhtajan, J. Sov. Math. {\bf 24}
(1984) 241.

\bibitem{LT} L.A. Takhtajan, Phys. Lett. {\bf A87} (1982) 479.


\bibitem{bado}
A. Doikou and A. Babichenko, Phys. Lett {\bf B515} (2001) 220; A.
Doikou, Nucl. Phys. {\bf B634} (2002) 591.

\bibitem{bydo} A. Bytsko and A. Doikou, in preparation.

\bibitem{BR} V.V. Bazhanov and N.Yu.
Reshetikhin, Int. J. Mod. Phys. {\bf A4} (1989) 115.

\bibitem{AM} S.R. Aladim and M.J.
Martins, J. Phys. {\bf A26} (1993) 7287.

\bibitem{fendleykink} P. Fendley, {\it cond--mat/9304031}

\bibitem{Bernard} D. Bernard, Phys.
Lett. {\bf B279} (1992) 78.

\bibitem{resrsos} N.Yu. Reshetikhin. J. Phys. {\bf A24} (1991) 3299.


\bibitem{abf}
G.E. Andrews, R.J. Baxter and P.J. Forrester, J. Stat. Phys. {\bf
35} (1984) 193.


\bibitem{rs} N.Yu Reshetikhin and H. Saleur, Nucl. Phys. {\bf B419} (1994)
507.

\bibitem{djmo}
E. Date, M. Jimbo, T. Miwa and M. Okado, Lett. Math. Phys. {\bf
12} (1986) 209.

\bibitem{nepo1}
R.I. Nepomechie, {\it hep--th/0211001}.

\bibitem{doikou-nepo} A. Doikou  and R.I. Nepomechie, J. Phys. {\bf A32} (1999) 3663.

\bibitem{rsosb1}
C. Ahn and W.M. Koo, {\it hep--th/9708080}; J. Phys. {\bf A29}
(1996) 5845.

\bibitem{rsosb2}
R.E. Behrend, P.A. Pearce and D.L. O'Brien, J. Stat. Phys. {\bf
84} (1996) 1.

\bibitem{rsosb3}
M.T. Batchelor, V. Fridkin, A. Kuniba and Y.K. Zhou, Phys. Lett
{\bf B735} (1996) 266.

\bibitem{bako} P. Baseilhac and K. Koizumi,  Nucl. Phys. {\bf B649} (2003) 491.

\bibitem{prep} P. Baseilhac, A. Doikou and K. Koizumi, in preparation.

\end{thebibliography}
\end{document}